\documentclass[preprint2]{aastex}

\newcommand{\etal}{et al.~}

\newcommand{\intfil}{Integral-shaped Filament}
\newcommand{\hii}{\ion{H}{2}}
\newcommand{\kms}{km s$^{-1}$}
\newcommand{\jybeam}{Jy beam$^{-1}$}

\shorttitle{Submillimeter Polarimetry of NGC 2068}
\shortauthors{Matthews \& Wilson}

\begin{document}

\title{Magnetic Fields in Star-Forming Molecular Clouds. IV. Polarimetry 
of the Filamentary NGC 2068 Cloud in Orion B}

\author{Brenda C.~Matthews\altaffilmark{1} \& Christine D.~Wilson}
\affil{McMaster University, 1280 Main Street West, Hamilton Ontario,
Canada L8S 4M1}
\altaffiltext{1}{Current Address: Department of Astronomy, 601 Campbell Hall, University of California Berkeley, Berkeley, CA 94720-3411}
\email{bmatthews@astron.berkeley.edu}
\email{wilson@physics.mcmaster.ca}

\begin{abstract}
We present submillimeter polarimetry at 850 \micron\ toward the
filamentary star-forming region associated with the reflection
nebulosity NGC 2068 in Orion B.  These data were obtained using the
James Clerk Maxwell Telescope's SCUBA polarimeter.  The polarization
pattern observed is not consistent with that expected for a field
geometry defined by a single mean field direction.  There are three
distinct distributions of polarization angle, which could represent
regions of differing inclination and/or field geometry within the
filamentary gas.  In general, the polarization pattern does not
correlate with the underlying total dust emission.  The presence of
varying inclinations against the plane of the sky is consistent with
the comparison of the 850 \micron\ continuum emission to the optical
emission from the Palomar Optical Sky Survey, which shows that the
western dust emission lies in the foreground of the optical nebula
while the eastern dust emission originates in the background.
Percentage polarizations are high, particularly toward the north-east
region of the cloud.  The mean polarization percentage in the region
is 5.0\% with a standard deviation of 3.1\%.  Depolarization toward
high intensities is identified in all parts of the filament.
\end{abstract}

\keywords{ISM: clouds, magnetic fields, molecules --- ISM: individual
(NGC 2068) --- polarization --- stars: formation --- submillimeter}

\section{Introduction}

The study of polarized emission from molecular clouds is of great
interest since, at long wavelengths ($\lambda > 25$ \micron), it
effectively traces the orientation on magnetic fields local to
star-forming regions \citep{hil88}.  Magnetic fields have been shown
to be energetically comparable to gravity and kinetic motions within
molecular clouds \citep{mg88,cru99,basu00} and are theorized to
provide vital support to clouds, preventing global collapse \citep[and
references therein]{mou87,mck93}.  Such support is necessary to
explain the low star-forming efficiencies observed in molecular
clouds, including Orion B, where the efficiency is $\sim 1$\%
\citep{car00}.

The Orion B cloud, at a distance of $\sim 415$ pc \citep{ant82}, is
the closest giant molecular cloud.  Within it, star formation is
concentrated into five distinct regions: NGC 2071, NGC 2068, LBS 23
(HH 24), NGC 2024 and NGC 2023, as determined from unbiased surveys
for young stellar objects \citep{lad91a} and dense CS gas
\citep{lad91b}.  Large scale 850 \micron\ dust emission from these
regions has been mapped by \citet{mit01} and \citet{mot01}.  Maps of
the polarized emission from NGC 2071, LBS 23, and NGC 2024 have
already appeared in the literature (see \citet[hereafter Paper
III]{mat01}).  In order to compare field geometries across Orion B, we
have now mapped the polarized emission from NGC 2068.  

The submillimeter emission from NGC 2068 lies south of the reflection
nebula (see Figure \ref{p4:pass+scuba}) which is seen optically and
contains an infrared cluster \citep{lad91a}.  The overall structure of
the dust emission is that of a ``clumpy filament'' in which 18
distinct compact sources are identified.  Most of the submillimeter
sources fall outside the boundary of the cluster identified in CS by
\citet{lad91b}.  Star formation is ongoing in this region, as
evidenced by detections of bipolar outflows by \citet{mit01} around
OriBN 51, from which evidence of outflow previously existed
\citep{edw84,gib00}, and also around OriBN 35 and OriBN 36 peaks (see
their Fig. 1c or Figure \ref{p4:map} below), although the sources of
these outflows are ambiguous due to the close proximity of OriBN 33
and OriBN 37 to the positions of high velocity gas.  OriBN 39 has a
2MASS infrared source associated with it, and thus should be a source
of outflow.  Evidence for redshifted gas near OriBN 47 also suggests
some outflow from this source.

There are no prior observations of polarized emission 
from the NGC 2068 dust emitting region.  However, \citet{man84}
measured linear polarization from scattering against the reflection
nebula north of the molecular condensations.  Based on the
centrosymmetric pattern observed, \citet{man84} ruled out the presence
of aligned grains within the reflection nebula.  They infer the
presence of a foreground assembly of grains illuminated from behind
solely by the star HD 38563N.  The position of this star is $\sim
5$\arcmin\ north-east of the dust emission on which we report in this
work, concident with the near-IR cluster observed by \citet{lad91a}.

Polarization data probe only two directions of the magnetic field
geometry -- those on the plane of the sky.  Additionally, they provide
no information about the strength of the magnetic field.  Where field
geometries are simple and the direction of the magnetic field does not
vary through the cloud depth, the polarized emission detected is
perpendicular to the mean magnetic field and the latter can be
inferred simply by rotating the polarization vectors by $90^\circ$.
If the field has a more complex, non-uniform geometry, then
interpretation becomes more difficult.  In such cases, it is best to
compare directly the polarization maps with polarization patterns
predicted from a physical model of a magnetized cloud.  It is
important to recognize that the polarizations measured are vector sums
along a particular line of sight through the cloud observed, weighted
by column density.  \citet{fp00a} present a model for a filamentary
cloud in which a helical magnetic field threads the filament and plays
an important role in determining the radial density structure.  This
model predicts an $r^{-2}$ density profile, which has been observed in
several clouds, including the \intfil\ \citep{joh99} and several
clouds in Cygnus \citep{lal99,all99}. The helical field geometry also
predicts depolarization toward the axis of a filament due to
cancellation effects on either side of the axis.  \citet{fp00c}
present predicted polarization patterns for cases in which the field
is either poloidally or toroidally dominated as well as for various
filament inclinations.  Qualitative extensions to these models have
been shown to reproduce observed polarization patterns in the
filamentary clouds OMC-3 \citep[hereafter Paper II]{mwf01} and NGC
2024 (see Paper III).  The filamentary structure of NGC 2068 is
therefore of particular interest as a further test of axially
symmetric magnetic field geometries.

This paper is the fourth in a series which seeks to compare the
polarization patterns (and inferred magnetic field geometries) in
different star-forming regions.  The observations and data reduction
techniques are described in $\S$\ref{p4:obs}.  The polarization data are
presented and analyzed in $\S$\ref{p4:poldata}.  The implications of
these data for the local magnetic field geometry and that of Orion B
as a whole are discussed in $\S$ \ref{p4:disc}, and $\S$ \ref{p4:summ}
summarizes our results.

\section{Observations and Data Reduction}
\label{p4:obs}

Using the UK/Japan polarimeter with the SCUBA (Submillimeter Common
User Bolometer Array) detector at the James Clerk Maxwell
Telescope\footnote{The JCMT is operated by the Joint Astronomy Centre
on behalf of the Particle Physics and Astronomy Research Council of
the UK, the Netherlands Organization for Scientific Research, and the
National Research Council of Canada.} (JCMT), we have mapped polarized
thermal emission from dust at 850 \micron\ toward a filamentary dust
cloud associated with the star-forming region NGC 2068.  The
observations were taken from 11 to 16 October 1999, and additional
data were added on 18 February 2000.  The polarizer and general
reduction techniques are described in \citet{gre01b} and
\citet{gre00}.  More information on data reduction and systematic
errors can be found in Paper II.
%\citet{mwf01}.

Six different SCUBA fields were required to map the entire NGC 2068
filament.  Initially, four fields were used, but then the number of
fields was increased and the positions shifted; hence 10 independent
field centers were used.  The field centers and other observing
parameters are summarized in Table \ref{p4:observing_parameters}.  The
data were obtained using a 16 point jiggle-map mode, in which the
telescope is ``jiggled'' in order to completely sample the SCUBA
field.  Chopping to a reference position was done to remove sky
effects.  The maximum chop throw possible is 180\arcsec.  We typically
used 150\arcsec\ to ensure we were not chopping onto polarized
emission.  The extinction of the atmosphere, $\tau$ (225 GHz), ranged
from 0.03 to 0.09 over the observations, but $>85$\% were taken when
tau(225 GHz) was 0.06 or 0.07.

We have reduced the data using the Starlink software package POLPACK,
designed to include polarization data obtained with bolometric arrays.
The data have been corrected for an error in the SCUBA computer's
clock which placed incorrect LST times in the data headers from July
1999 to May 2000.  This error did not affect the telescope's
acquisition or tracking but affects data reduction since the elevation
and sky rotation are calculated from the LST times in the data files.
The evolution of the magnitude of this error over time has been
tracked and can therefore be corrected retroactively by adjusting the
times in the headers (see the JCMT website for details).  The error in
timing after this adjustment is $\pm 10$s.

After extinction correction, noisy bolometers were identified and
removed from the data sets.  The data were sky subtracted using
bolometers with mean values close to zero, but not those which were
significantly negative.  (The area mapped by each bolometer in a
single jiggle map can be identified with reasonably accuracy, since
little sky rotation occurs during the 16-point jiggle.  Hence, by
examining the flux levels of pixels within a bolometer, those in which
a majority of adjacent pixels are negative were considered
significantly negative and not used.  Those for which negative and
positive values seem equally distributed can be considered to contain
zero flux and be used for sky subtraction.)  At 850 \micron, the sky
is highly variable on timescales of seconds.  This variability must be
measured and removed from the data.  Chopping removes the effects of
slow sky variability; however, fast variations remain in the data,
which require sky subtraction using array bolometers devoid of
significant flux.  We used between one and three bolometers to
determine the sky variability, using the existing scan maps of total
intensity at 850 \micron\ \citep{mit01} to help select empty
bolometers.  The methods of sky subtraction are discussed in detail in
\citet{jen98}.  As part of our analyses of other regions, we
determined that sky subtraction can be done effectively with a single
bolometer (see Paper II)
%\citep{mwf01} 
and that the POLPACK reduction process is
extremely robust with respect to the selection of sky bolometers
(see Paper III).
%\citep{mat01}.  
We have not added the mean flux removed by sky
subtraction back into the NGC 2068 maps, since the flux in the sky
positions was sufficiently close to zero (as determined from the large
scale intensity map of \citet{mit01}).

Once the instrumental polarizations (IPs) were removed from each
bolometer, all the data sets were combined to produce a final map in
three Stokes' parameters, $I$, $Q$ and $U$, where $I$ is the total,
unpolarized intensity and $Q$ and $U$ are two orthogonal components of
linearly polarized light.  These three Stokes' parameters were then
combined to yield the polarization percentage, $p$, and polarization
position angle, $\theta$, in the map according to the relations:
\begin{equation} 
p = \frac{\sqrt{Q^2 + U^2}}{I}; \ \ \  \theta = \frac{1}{2} \arctan(U/Q).
\label{p4:values}
\end{equation}

\noindent The uncertainties in each of these quantities are given by:
\begin{equation}
dp= p^{-1}\sqrt{[dQ^2Q^2 + dU^2U^2]}; \ \ \ d\theta = 28.6^\circ / \sigma_p
\label{p4:uncertainties}
\end{equation}

\noindent where $\sigma_p$ is the signal-to-noise in $p$, or $p/dp$.

\noindent A bias exists which tends to increase the $p$ value, even
when $Q$ and $U$ are consistent with $p=0$, because $p$ is forced to be
positive.  The polarization percentages were debiased according to the
expression:
\begin{equation}
p_{db}= \sqrt{p^2-dp^2}.
\label{p4:debias}
\end{equation}

\noindent Future references to polarization percentage, or $p$, refer
to the debiased value.

In order to improve the $\sigma_p$, and hence $d\theta$, the data were
binned to 15\arcsec\ sampling.  This rebinning improves $\sigma_p$ by
a factor of five over the unbinned 3\arcsec\ sampled data.  Good
polarization vectors were selected to have $\sigma_p > 3$ (which
implies $d\theta < 10^\circ$), $dp<1$\%, $p>1$\% and be coincident
with positions where the total unpolarized 850 \micron\ flux exceeds
20\% of the faintest of the six compact peaks, OriBN 35.  The
filtering of polarizations less than 1\% accounts for the
uncertainties in the IP values, as well as for any contamination due
to sidelobe polarization, as discussed in \citet{gre01b}.  Sidelobe
contamination refers to a polarized signal measured within the main
beam due to a source within a sidelobe.  If there is enough polarized
flux ($p \times I$) from a source in a sidelobe position, a signal can
be produced in the main beam despite the fact that the power in the
JCMT sidelobes at 850 \micron\ is typically $\le 1$\% that of the main
beam.  \citet{gre01b} derive the minimum believable polarization
percentage, $p_{crit}$, based on: the ratio of power in the sidelobe
to the main beam, $P_{sl}/P_{mb}$; the ratio of unpolarized, total
flux of the source in the sidelobe to that in the main beam,
$F_{sl}/F$; and the IP estimate at the sidelobe, $p_{sl}$:
\begin{equation}
p_{crit} \ge 2 \times p_{sl} \left (\frac{P_{sl}}{P_{mb}} \right )
\left (\frac{F_{sl}}{F} \right ) .
\label{p4:pcrit}
\end{equation}

We can estimate the extreme value of $p_{crit}$ for the entire NGC
2068 region by calculating the worst case scenario.  This occurs for the
field centered at $\alpha (\rm{J2000}) = 05^{\rm h}46^{\rm m}$31\fs2
and $\delta (\rm{J2000}) = -$00\degr00\arcmin25\farcs5, in which the
brightest source in our map, OriBN 51, was approximately 54\arcsec\
from the center of the array.  The beam responses and IP values are
obtained from maps of unpolarized planets.  Using polarization maps of
Saturn from 14 October 1999, 16 October 1999 and 18 February 2000, the
ratios of power in the sidelobe versus main beam are deduced to be
0.009, 0.01, and 0.004 respectively.  The corresponding mean IP values
at this sidelobe position are 4.5\%, 4.7\%, and 4.3\%.  The ratio of
the flux of Ori BN 51 to the center of the field is $\sim 7$.  Hence,
using equation (\ref{p4:pcrit}), the $p_{crit}$ values are determined to
be 0.57\%, 0.66\%, and 0.24\%.  Thus, by selecting only values of $p >
1$\%, the polarizations cannot be attributable to sidelobe
contamination.

\section{NGC 2068 Polarization Data}
\label{p4:poldata}

Figure \ref{p4:map} shows the polarization data toward the NGC 2068
filamentary cloud.  These data are displayed over a portion of the
\citet{mit01} map of Orion B North.  The polarization data are binned
to 15\arcsec, slightly greater than the 14 \arcsec\ beamwidth of the
JCMT at 850 \micron.  Thus, each vector can be treated as an
independent measurement of the local polarization at 850 \micron.  The
data are presented in tabular form in Appendix \ref{appendixA}.  By
eye, one can readily discern that three different populations of
position angles exist within Figure \ref{p4:map}.  Separation of the
data into three regions based on boundaries in RA can generally
separate these populations.  Figure \ref{p4:PAhist} shows the
distributions in position angle over the whole map, and for each of
the three subsections.  Region 1 encompasses the eastern part of the
map, including all vectors east of the J2000 position of $05^{\rm
h}46^{\rm m}$39\fs8.  Region 2 is west of Region 1, extending to RA
$05^{\rm h}46^{\rm m}$33\fs8 (J2000).  Region 3 covers the western
section of the map, including all vectors west of $05^{\rm h}46^{\rm
m}$33\fs8 (J2000).

NGC 2068 contains six compact dust condensations \citep{mit01}, as
well as a dozen more amorphous, fainter condensations.  In the OMC-3
filament of Orion A, the polarization pattern was no different in the
presence of embedded cores than elsewhere along the filament
(Paper II).
%\citep{mwf01}.  
However, in Figure \ref{p4:map}, polarization vectors near cores are
smaller and more variable in direction, particularly near the cores
OriBN 35, 47 and especially 51.  In the latter case, no polarization
is detected toward the core.  Toward each of these cores, there is
evidence of outflow; they are not pre-protostellar in nature.  We note
however that across the outflow source OriBN 39 and the compact core
OriBN 38, the polarization vectors are consistent with the patterns
surrounding the cores.  The effect of the cores on the polarization
pattern indicates that the filament is not dominating the polarized
emission as was the case in OMC-3 in Orion A (see \citet{mat00} and
Paper II).
%\citep{mwf01}.

\subsection{Polarization Position Angle and Filament Orientation}

In Region 1, the mean polarization position angle is $109^\circ$ east
of north, which is equivalent to $-71^\circ$ in linear polarization
(since vectors offset by $180^\circ$ are indistinguishable).  The
amorphous cores near the north-east have the highest polarizations,
and the vector orientations there are roughly perpendicular to the
projected filamentary axis, aligned in a roughly north-east to
south-west orientation. Interestingly, the vectors along the three
cores OriBN 34, 35 and 39 align well with those to the north-east, but
now appear to lie parallel to the string of three cores.  In Region 2,
which includes the cores OriBN 37, 38 and part of 36, the vectors also
follow the string of cores.  Finally, in Region 3, which includes all
cores west of OriBN 42, the polarization vectors are distributed about
a mean of $74^\circ$, although the projected filament direction
varies considerably.  The vectors thus tend to follow the filament
near OriBN 42 to 48, and then lie perpendicular to the filamentary
dust emission along the whole eastern edge.  Overall, there is no
strong correlation between the filament orientation and the vector
position angles.

Table \ref{p4:poltable} summarizes the statistics of the three data
subsets plus the whole data set for the NGC 2068 region, as well as
the results of single Gaussian fits to each of the position angle
distributions shown in Figure \ref{p4:PAhist}.  The mean position
angle is denoted $\mu$, while the width of the distribution is
$w(\theta)$.  The reduced chi-squared value is also listed and in each
case is $<1$.  The fit to Region 3 gives a mean of $66^\circ$ with a
width of $18^\circ$.  Region 2 gives a mean of $93^\circ$ with a width
of $16^\circ$, and Region 1 yields a mean of 123$^\circ$ and
width of $24^\circ$.  The results of the fits are consistent with the
statistical mean and standard deviation in the position angles.

\subsection{Polarization Percentage}
 
The results of a basic statistical analysis of the polarization
percentage distributions in the NGC 2068 cloud and its subset of three
regions are reported in Table \ref{p4:poltable}.  The mean polarization
percentage in the whole filamentary cloud is 5.0\% with a standard
deviation of 3.1\%.  Similar values are found for each separate region
within NGC 2068.  No systematic variation in polarization percentage
is obvious from examination of Table \ref{p4:poltable}.  Interestingly,
Region 2 has the highest median value of $p$, and it is the only
region where the median exceeds the mean, which indicates that in this
region, polarization is high.  The mean is not biased by only a few
vectors of particularly high polarization percentage.

One of the most interesting properties of dust emission polarimetry
has been the measurement of declining polarization percentage with
increasing unpolarized intensity.  Figure \ref{p4:depol} shows this
same trend in the NGC 2068 region and its subregions that has been
observed in other regions.  Over the entire data set, the variation of
$p$ with total intensity follows a power law of the form: $p = A
\times I^\gamma$, where $\gamma = -0.81 \pm 0.04$ and $A = (1.6 \pm 
0.4) \times 10^{-2}$.  No systematic variation in polarization
percentage between regions is evident from examination of Figure
\ref{p4:depol}.  We do note, however, that the effect is strongest in
Region 3.  In Region 1, the vectors surrounding the cores OriBN 31 and
32 are shown as filled triangles, while the rest of the region's
vectors are plotted as open triangles.  It is clear that the highest
polarizations are observed at the extreme north-east of the NGC 2068
region.

As discussed in Paper II,
%\citet{mwf01}, 
a depolarization effect can be produced
systematically by chopping the telescope during observing.  However,
based on that analysis, the steep profiles shown in Figure \ref{p4:depol}
could not be explained by this effect unless there were {\it
significant} polarized flux in the chop position.  For example, if the
reference, or chop, position had 25\% of the peak flux in the source
field and was polarized to twice the degree of the source field, then
a slope of $-0.88$ could be produced on a log $p$ versus log $I$ plot
{\it in regions of low total intensity}.  Even under this extreme
scenario, in regions of high flux, we would not expect such a sharp
decline in polarization percentage with intensity.  We thus conclude
that the depolarization effect in this region is not produced by
systematic effects of chopping.

\section{Discussion}
\label{p4:disc}

\subsection{Polarization Percentage in Orion B}

Figure \ref{p4:Pcompare} shows the distributions of polarization
percentage in the regions of Orion B observed with the SCUBA
polarimeter.  These distributions have been normalized to the total
number of vectors measured in each region, and the counts expressed as
a fraction of the total.  The error bars represent the $\sqrt{N}$
statistical uncertainty in the number of counts, also normalized to
the total in each map.  The distribution for NGC 2068 is comparable to
those measured in the other star-forming regions of Orion B North --
NGC 2071 and LBS23N -- and with measurements toward the OMC-3 filament
in Orion A (Paper II).
%\citep{mwf01}.  
The NGC 2024 region is dominated by low percentage polarizations, with
the majority between 1-2\%.  This result could imply NGC 2024 has
weaker magnetic fields, poorer grain rotation or alignment, a
different grain composition, or some combination of these factors.
The fact that NGC 2071, NGC 2068 and LBS 23N show distributions which
are reasonably flat from 1-6\% could indicate similar grain
properties, field strengths, and degrees of grain aligment, but this
cannot be proven with these data.  The densities toward the NGC 2024
cores have been estimated to be abnormally high ($10^8$ cm$^{-3}$
\citet{mez88}), but \citet{sch91} suggest the values are more typical
of cores ($10^6$ cm$^{-3}$).  Polarization data at other wavelengths
(such as 350 \micron) could help constrain the dust properties
according to models (see \citet{hil00} and references therein), and
observations of dense molecular tracers like OH for Zeeman splitting
could provide more detailed information about the field geometries in
these four regions.

Toward all four regions of Orion B observed in polarized emission, the
depolarization effect is detected (see Fig.\ 5 of Paper III
%\citet{mat01} 
and Figure \ref{p4:depol}).  Paper III
%\citet{mat01} 
shows that NGC 2024 exhibits a similar depolarization signature to the
northern core NGC 2071.  However, LBS 23N has a significantly steeper
slope of $-0.95$, which is more consistent with that of Region 3 in
our data set.  This trend has been observed in many other regions as
well, including massive cores such as OMC-1 \citep{sch97} and
protostellar and starless cores \citep{gir99,war00}.

\subsection{Evidence for Varying Inclinations in NGC 2068}

\begin{figure}
\vspace*{12cm}
\includegraphics{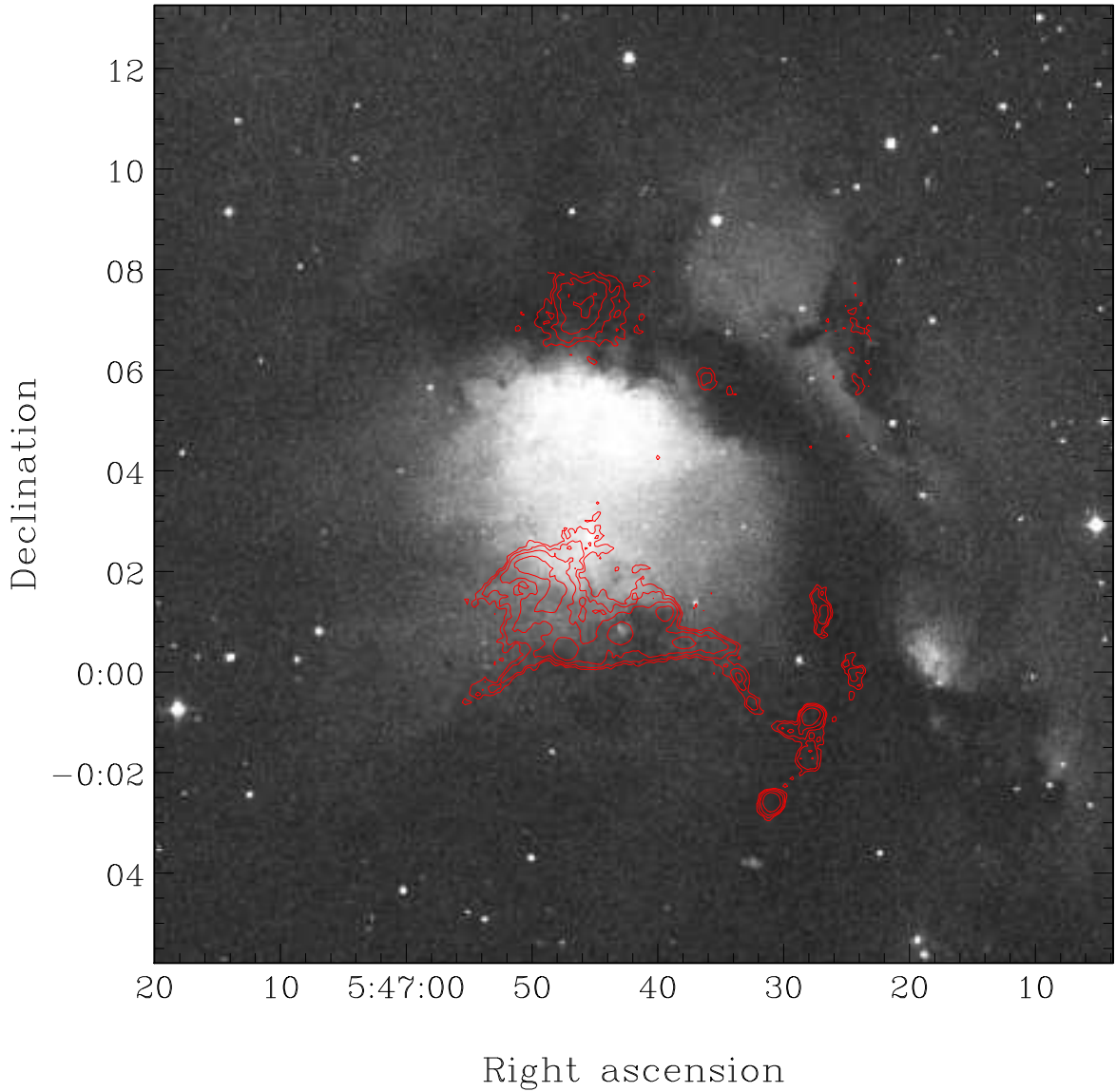}
\caption{The optical and dust emission
from NGC 2068. The optical image from the Palomar Sky Survey is
presented, covering an area of $\sim 20$\arcmin $\times \sim
20$\arcmin.  The NGC 2068 reflection nebula is central.  The SCUBA
total, unpolarized dust emission at 850 \micron\ is shown in contours
of 90, 93, 96 and 99 percentiles from \citet{mit01}.  The western edge
of the dust emission appears to be coincident with the optical dust
lane.  At the east, however, the region of dust emission appears
bright in the optical, and hence the dust must lie behind the nebula.
A northern dust peak has not yet been observed with the polarimeter.
}
\label{p4:pass+scuba}
\end{figure}

The filamentary dust emission of NGC 2068 likely arises from dust at
different depths in the Orion B cloud.  The comparison between the
Palomar Observatory Sky Survey optical data and the dust emission has
been made by \citet{mit01}. We produce a similar image in Figure
\ref{p4:pass+scuba}, which shows that while the western dust emission
lines up well with the optically dark dust lane, the eastern dust
emission has no corresponding dark lane.  (North of the \hii\ region
lies another dust condensation not yet observed with the polarimeter.)
Hence, it is likely that the dust emission may lie on the outer edge
of the reflection nebula with the western material in the foreground
and the eastern material behind.  Thus, it is clear that this filament
does not lie in the plane of the sky, and that the inclination on the
sky is likely variable.

An obvious question is whether or not all the dust emission arises
from material which is spatially related.  It is noteworthy that the
core OriBN 34 does not appear to line up well with the filamentary
material of Figure \ref{p4:map}.  In fact, our comparison of the
optical and dust emission shows that the location of OriBN 34 is
optically dark.  This could be a coincidence, or OriBN 34 could be
closer than the dust emission appearing next to it on the SCUBA image.
OriBN 39 also appears dark and could be in the foreground.
\citet{mit01} present $^{13}$CO $J=2-1$ toward NGC 2068 and a partial
map of the north-eastern region in C$^{18}$O $J=2-1$ (see their
Figs. 6 and 7).  The $^{13}$CO contours are closely correlated to the
850 \micron\ dust emission.  The same is true for the C$^{18}$O
emission where data exist.  These maps are integrated over velocity
ranges from 5 \kms\ $< v_{LSR} <$ 15 \kms\ and 7 \kms\ $< v_{LSR} <$
13 \kms, respectively.  Thus, the emission is confined to the Orion B
cloud.  The fact that OriBN 34 and 39 show similar polarization
position angles as the eastern area but may be in the foreground would
argue against these regions being completely spatially distinct
(unless the cores are not contributing as much polarized emission as
the background diffuse material.  If the material is spatially
separated, then the similar orientations of the polarization position
angles could suggest that the field, if not defined by a single mean
field direction, could be organized on spatial scales at least as
large as the separation between them.

The 2.4 \micron\ emission from the NGC 2068 reflection nebula spans
approximately 6\arcmin\ in spatial extent \citep{sel84}.  This is
consistent with the physical size of the nebula in optical emission as
shown in Figure \ref{p4:pass+scuba}, and at a distance of 415 pc
corresponds to 0.7 pc.  If the nebula is as deep as it is wide, then
the source of dust emission at the north-east could be more than a
parsec displaced spatially from the cores at the west (taking
7\arcmin\ as an estimate of their separation in projection).

\subsection{Field Geometry}

\subsubsection{A Single Mean Field Direction}

The vectors of NGC 2068 do not support a single field orientation in
NGC 2068.  If one assumes that the magnetic field geometry throughout
a cloud's depth is reasonably well defined by a mean field direction,
then the direction of the field can be obtained by rotating emission
polarization data by $90^\circ$.  Doing this in the NGC 2068
polarization map will clearly not produce aligned vectors, any more
than the polarization data themselves are aligned.  Therefore, we can
rule out a uniform, unidirectional field across all of NGC 2068.  This
was also the case in OMC-3 in Orion A (see Paper II); however, in that
region, the filamentary axis was easy to define, and a strong
correlation between the axis and polarization position angle was
observed along 75\% of the filament.  The vectors of NGC 2068 do not
show an obvious alignment of polarization position angle with the
filament orientation.

We can compare the polarization position angles across those regions
observed thus far in Orion B.  In the three regions of Orion B North,
no evidence for similar polarization orientations exists.  Toward NGC
2071, the polarization vectors align generally with the prominent
outflow from the IRS 1 source at a position angle of $40^\circ$ east
of north.  Within the LBS 23N string of cores to the south of NGC
2068, the vectors are generally aligned north-south (position angle
$0^\circ$) although the scatter in this faint region is considerable.
Neither of these orientations is dominant in NGC 2068.  Thus, there is
no support for a mean field direction in NGC 2068 or across the three
star-forming regions of Orion B North.

The one region observed in Orion B South is NGC 2024, and the
polarization pattern from that region has been modeled as arising
either from a helical field geometry or from the expansion of the
ionization front due to the associated \hii\ region in Paper III.
%\citet{mat01}.  
The latter geometry is favored since it is most compatible with the
total physics and geometry of NGC 2024.  It is clear that the
polarization patterns across these star-forming regions can be
strongly correlated with the dust and gas structures of a particular
region, and that each region must be modeled separately.

\subsubsection{More Complex Geometries}

As discussed above, there is evidence that the dust arises from a
connected gas structure.  However, the spatial separation between the
filament edges could be $> 1$ pc.  The regions of different position
angles could indicate regions of different field geometry or
inclinations.  We propose two model geometries which could potentially
explain the variable position angles observed.

Since the cores are mainly aligned along a filamentary structure, a
helical field geometry is appealing.  It can explain the confinement
of gas in the filament, the fragmentation to cores and the elongation
of the asymmetric concentrations along the axis of the filament, which
is certainly the case for OriBN 31, 32, 33, 41 and 49.  \citet{fp00c}
show possible polarization patterns for different helical field
conditions (i.e.\ poloidally-dominated versus toroidally-dominated).
These models are developed for straight filaments, and \citet{fp00c}
find that for such filaments, the polarization vectors align either
parallel or perpendicular to the projected filament axis regardless of
the inclination of the filament.  However, Paper II
%\citet{mwf01} 
presents a
model for a helical field threading a bent filament.  In this case,
the vectors may adopt any orientation relative to the filament due to
the asymmetries in the filament.  The relative alignment depends on
the filament's inclination and rotation on the plane of the sky.

Thus, as inclination and angle in the plane of the sky vary, a
helically-threaded filament should produce different polarization
position angles relative to the projected filament orientation.  For
instance, if the polarization positions are aligned with the filament,
a toroidally-dominated helical field could wrap the filament locally.
Conversely, where the vectors appear perpendicular to the filament, a
helical field would be poloidally-dominated, or the filament and field
could be significantly bent.  At the north-western edge near the cores
OriBN 33 and 41, the vectors are neither parallel nor perpendicular to
the filament.  This could also indicate that the filament is bent
along its length in this area.  Modeling such a complex filamentary
structure will be difficult, since more than one filament/field
geometry can produce similar polarization patterns.

One additional test for the presence of a helical field geometry is
the radial profile of the total, unpolarized emission.  Using the 850
\micron\ map of \citet{mit01}, a radial profile can be built up by
taking several slices across the filament.  We have attempted to
confine the cuts to the regions between cores along NGC 2068.  Figure
\ref{p4:slice} shows the profiles through several such slices, taken
between the OriBN 44 and 48, between 32 and 39 and between 35 and 36.
For the second cut, both sides of the profile were used; for the other
two, one side of the slice was discarded due to the presence of
extended emission from the filament near OriBN 47 (for the cut between
OriBN 44 and 48) and extended emission from core OriBN 34 (for the cut
between OriBN35 and 36).  The fluxes have been normalized to the
maximum through the cuts (respectively 0.07 \jybeam, 0.32 \jybeam, and
0.28 \jybeam); this position is assumed to be the axis of the
filament.  The NGC 2068 filament is so narrow and faint, it is
difficult to interpret the profiles.  To guide the eye, we have drawn
on Figure \ref{p4:slice} lines corresponding to slopes of -0.5 and -1.
Near the axis, the flux falls off with a distribution consistent with
a power law index of $-0.5$.  Toward lower fluxes, the index could be
closer to $-1$; however, at these low flux levels, interpretation of
the index becomes difficult.  The flux profile of $r^{-0.5}$
corresponds to a density profile of $r^{-1.5}$ for an isothermal
filament.  The helical field geometry predicts a density profile with
index $-2$ for an isothermal equation of state.  For other equations
of state (e.g.\ the logotrope, \citet{mcl96}), a shallower range of
density profiles can be generated.  Unmagnetized, isothermal filaments
predict much steeper profiles, with indices of $-4$, which are
clearly not consistent even with these poor profiles.

A second possible field geometry is suggested by the position of the
molecular filament so near the periphery of the reflection nebula of
NGC 2068.  Paper III
%\citet{mat01} 
presents a model of the polarization data
toward the NGC 2024 dense ridge of cores in which the field,
compressed by the expansion of the NGC 2024 \hii\ region, is then
moulded around the dense ridge as the ionization front approaches the
cores.  This picture accounts for both the polarization pattern at 850
\micron\ and the measurements of the line-of-sight field direction
and strength measured by \citet{cru99}.  The expansion of the
reflection nebula could produce a similar effect in NGC 2068, although
the pattern produced is much more complex.  In NGC 2024, the ridge is
entirely located on the far-side of the \hii\ region, whereas in NGC
2068 the comparison of optical and 850 \micron\ dust emission in
Figure \ref{p4:pass+scuba} indicates that the dust emission arises from
both the near and far-side of the nebula.

A recent publication on the formation of quiescent cores through
turbulent flows \citep{pad01} provides a third possible
interpretation of the polarization pattern.  This work discusses the
potential formation of cores due to the presence of super-sonic
turbulent flows within molecular clouds.  In this model, cores form by
accretion along filamentary structures, with the brightest cores
forming at the loci of intersecting shocks.  The core OriBN 47 lies at
the intersection of three filamentary segments.  As predicted by
\citet{pad01}, this core exhibits depolarization, although at the
distance of Orion B, it is difficult to achieve a good sampling of
polarization vectors across this core.  Thus, we do not see the large
changes in position angle predicted near cores (see their Fig.\ 3) and
routinely observed in Bok globules and starless cores in closer
regions of star formation (e.g.\ see \citet{war00}).  However,
the polarization pattern along the larger scale filamentary structure
is well sampled in our map.  In the turbulent flows model, the
polarization vectors along the filamentary structures are seen to
align well with the filaments' axes.  In NGC 2068, only the filament
segment east of OriBN 48 exhibits this behavior.  The segments to the
north-west and south-west show vectors oriented roughly perpendicular
to the filamentary structure, which does not agree well with the
turbulent flows picture.  A simulation of turbulent flows along
filaments of higher density may agree better with our
observations in NGC 2068.  If a threshold in extinction exists beyond
which grains are not effectively aligned \citep{pad01}, then
denser turbulent flows could exhibit less correlation between the
filamentary axis and the inferred polarization.

\section{Summary}
\label{p4:summ}

We have detected polarized emission from aligned dust grains at 850
\micron\ with the SCUBA polarimeter.  These data reveal strong degrees
of polarization (up to 17\%) toward the NGC 2068 region, with higher
polarization percentages detected toward regions of fainter total
intensity.  This depolarization effect has been observed in most
regions of polarized emission.  Significant depolarization has been
detected toward several compact cores along the filament.  Within the
cores of LBS 23N and some cores along the NGC 2024 dense ridge,
significant depolarization was also measured (see Paper III)
%\citep{mat01}.  
We note that this was not the case in the OMC-3 filament of Orion A,
where the polarization patterns behaved consistently in the presence
or absence of embedded cores (see Paper I and Paper II).
%\citep{mwf01}.  
This was attributed to the fact that
filamentary dust dominated the polarization pattern in Orion A,
but that this is not the case in Orion B.

The polarization pattern is inconsistent with that expected for a
uniform field geometry, for which vectors should be aligned with one
another across the entire region.  The polarization position angles
are roughly organized into three distinct distributions which do not
align with the orientations of the filamentary total intensity
emission.  This is also different than the pattern observed in the
OMC-3 part of the \intfil\ in Paper II
%\citet{mwf01} 
where the vectors aligned with the
filament orientation along 75\% of the filament's length.  Taken in
comparison with the polarization data for three other star-forming
regions in Orion B (Paper III),
%\citep{mat01}, 
there is also no evidence for a
single field orientation across the Orion B cloud as a whole.

Comparison of optical and 850 \micron\ dust emission establishes that
the dust filament is not lying in the plane of the sky, but must be
twisted such that its western edge is in the foreground of the optical
reflection nebula, while the eastern edge lies behind.  These changes
in inclination can help explain the polarization pattern observed.
Paper II
%\citet{mwf01} 
shows that a bent filament threaded by a helical field
geometry can produce polarization vectors offset from the filament
orientations by varying degrees.  Material lying on the edge of an
expanding reflection nebula could also exhibit different magnetic field
orientations at different locations depending on how effectively the
field has been compressed by the expansion.  The complexity of this
filamentary structure will make modeling difficult, since there may
be many degenerate choices of filament/field configurations which can
produce the observed polarization pattern.

These are the first observations of polarized emission toward this
filament.  The strength of the polarization percentages detected
suggests that polarized emission should be detectable at other
wavelengths.  For instance, observations at 350 \micron\ with the
Hertz polarimeter would provide some information on the polarization
spectrum for this region.  This could lead to constraints on the grain
population along the filament.  Additionally, Zeeman data on $B_{los}$
field would provide significant constraint to the three-dimensional
field structure.

\acknowledgements

The authors thank the members of the Canadian Consortium on Star
Formation for the 850 \micron\ scan map of the Orion B North cloud.
Thanks to J.~Greaves, T.~Jenness, and G.~Moriarty-Schieven at the JCMT
for assistance during and after observing.  The research of BCM and
CDW has been supported through grants from the Natural Sciences and
Engineering Research Council of Canada.  BCM acknowledges funding from
Ontario Graduate Scholarships during the period of this research.

\appendix
\section{The Polarization Data}
\label{appendixA}

\begin{figure}
\vspace*{12cm}
\includegraphics{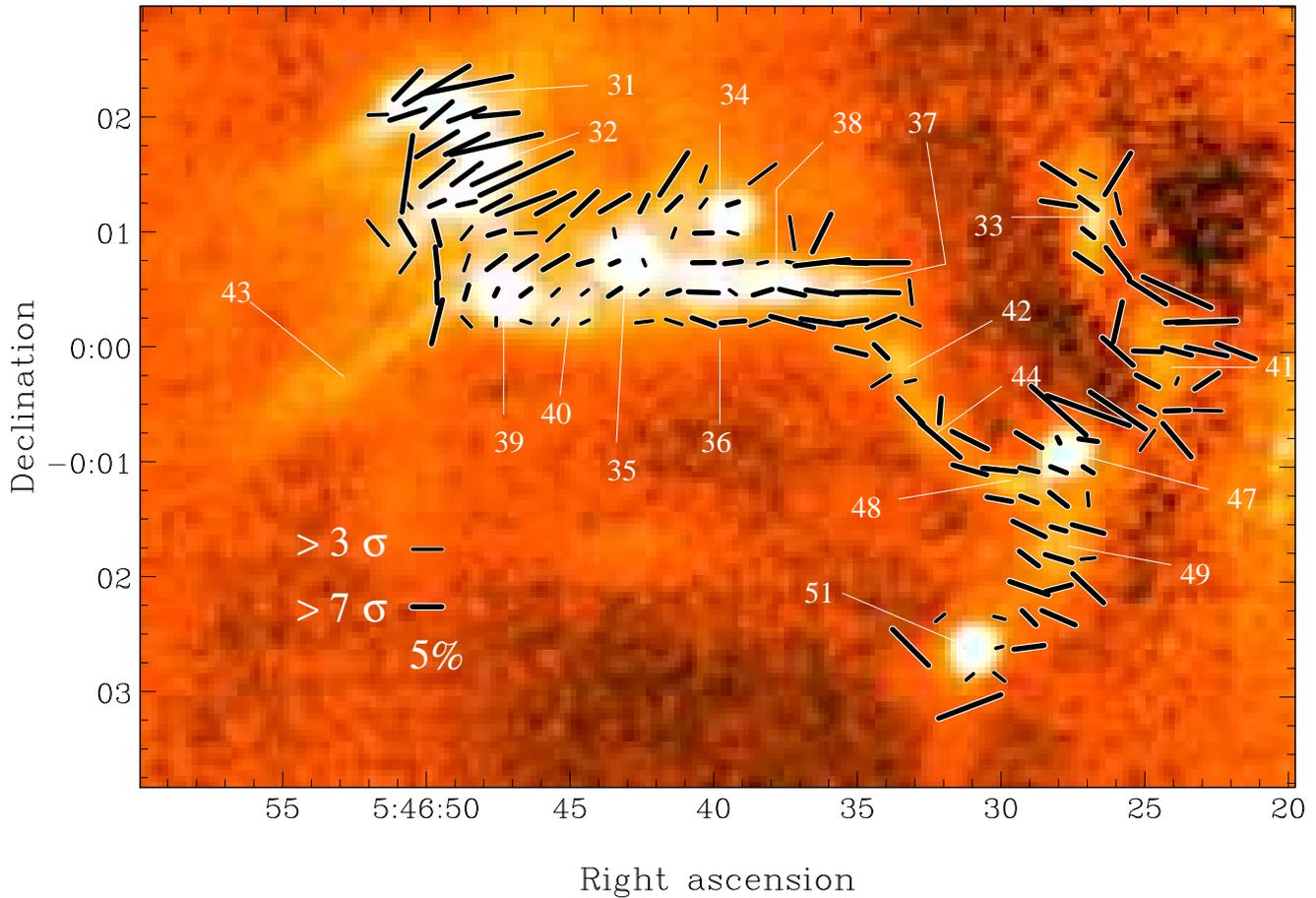}
\caption{The 850 \micron\ polarization pattern across NGC
2068. The colored greyscale is a more spatially extensive scan map
provided by \citet{mit01}.  The greyscale range is $-2\sigma$ to
$3\sigma$.  Polarization data were sampled at 3\arcsec\ and have been
binned to 15\arcsec\ ($>$ the JCMT beamwidth).  All vectors plotted
are coincident with location where the total, unpolarized intensity $>
20$\% that of OriBN 35 (10\% that of OriBN 51).  The thin vectors have
signal-to-noise $\sigma_p > 3$, while the thick vectors have $\sigma_p
> 7$.  The thin vectors are accurate in position angle to better than
10$^\circ$, while the thicker vectors are accurate to better than
4$^\circ$.  The coordinates of the map are J2000.}
\label{p4:map}
\end{figure}

\begin{figure}
\vspace*{15cm}
\includegraphics{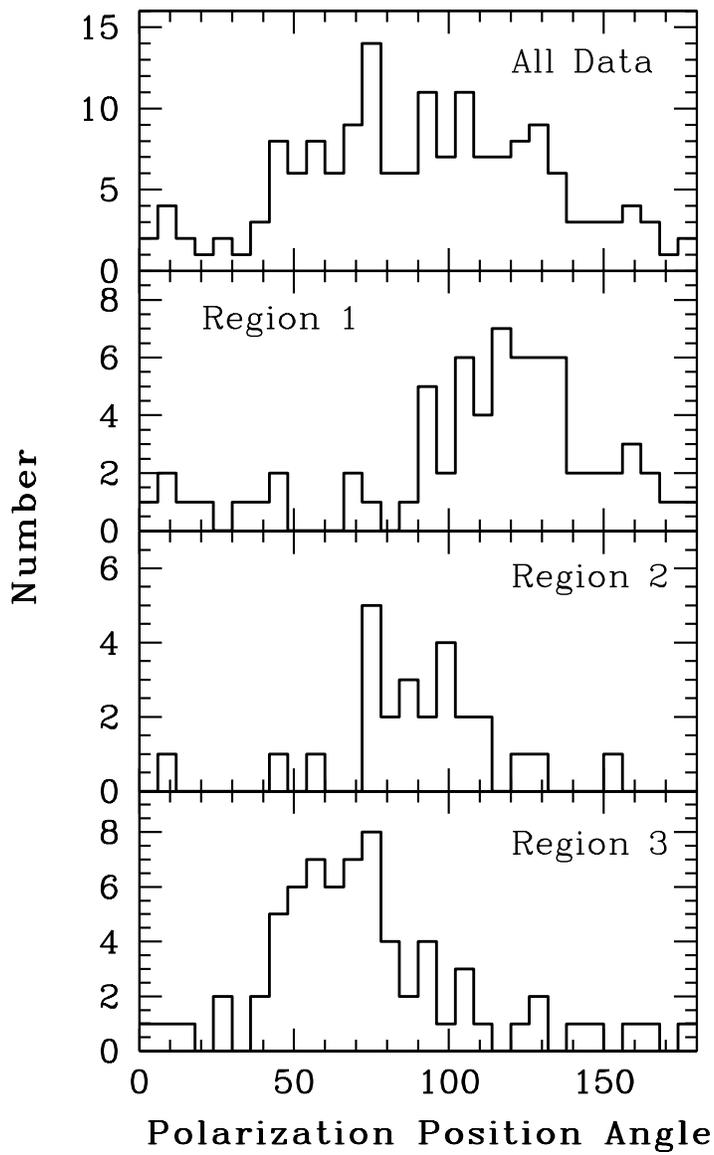}
\caption{The distribution of polarization position angle
in NGC 2068. The position angles of all vectors plotted on Figure
\ref{p4:map} are presented in the top panel.  The distribution of
vectors has a mean of $91.4^\circ$ and standard deviation about the
mean of $40^\circ$.  Additionally, the distribution of position angles
in all three subsets of the data as defined in the text are shown.}
\label{p4:PAhist}
\end{figure}

\begin{figure}
\vspace*{15cm}
\includegraphics{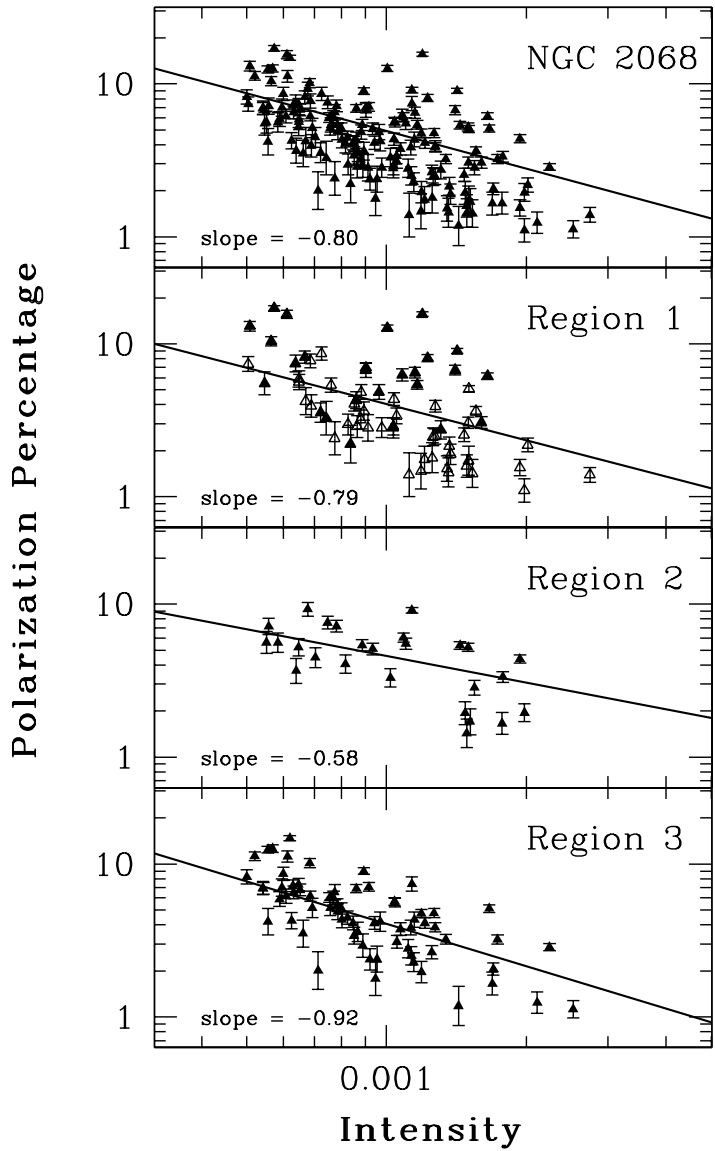}
\caption{The depolarization effect in NGC
2068.  We have plotted $p$ versus $I$ in log space and included fits
to the corresponding $p$ versus $I$ profiles.  The fits are to the
power law: $p = A \times I^\gamma$, where $\gamma$ is shown as the
slope of the log-log plot.  Plots are shown for each of the subregions
discussed in the text.  The Region 1 panel illustrates positions at
the extreme north-east (near OriBN 31 and 32) as filled triangles,
while the rest of the vectors in the area are plotted as open
triangles.  The strongest polarizations are located in the north-east
area.}
\label{p4:depol}
\end{figure}

\begin{figure}
\vspace*{15cm}
\includegraphics{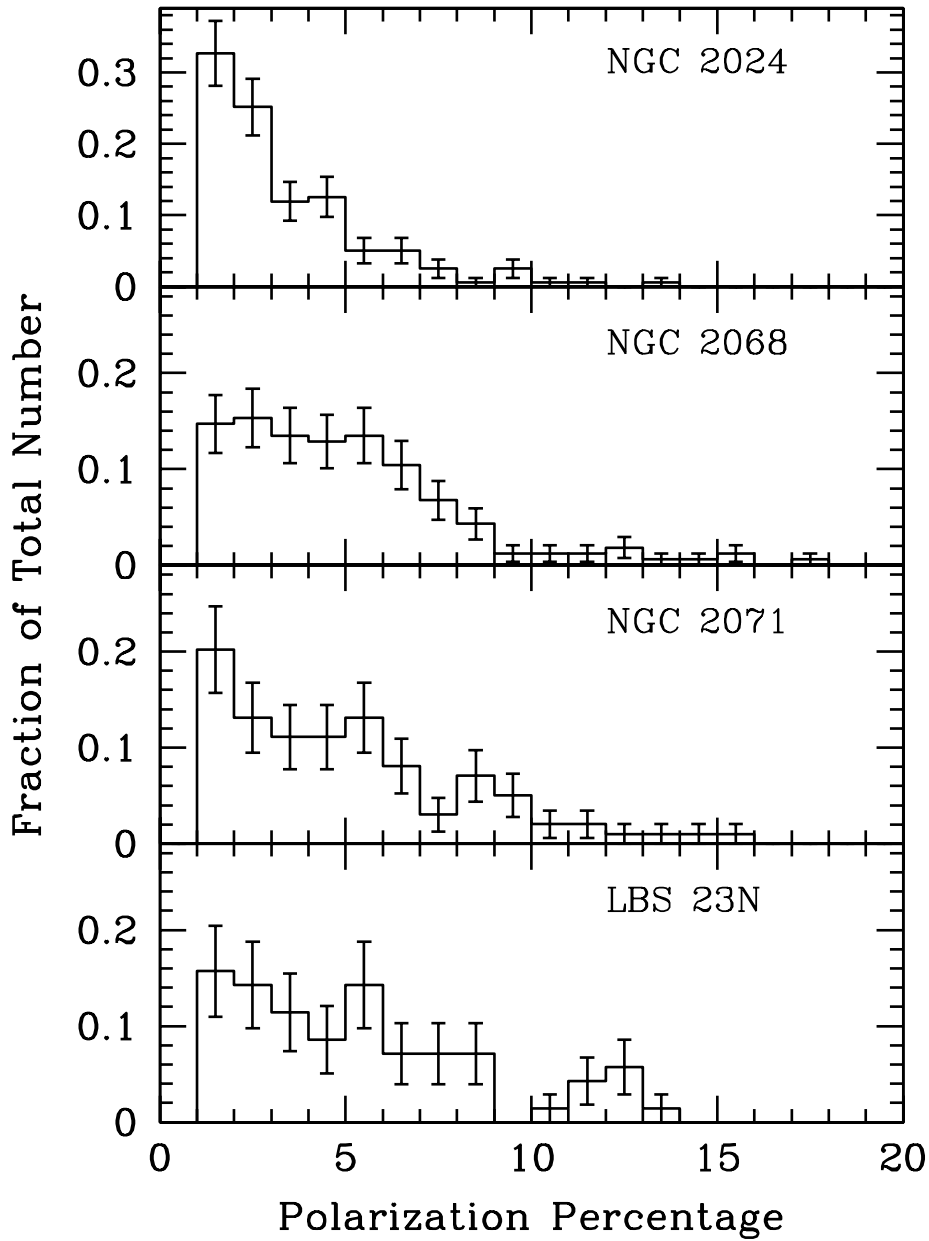}
\caption{The distribution of polarization
percentage in four regions of Orion B. The distributions for NGC 2024,
NGC 2071 and LBS 23N are from data presented in Paper III.
Clearly, the distribution of NGC 2024 is most heavily weighted to
small values of $p$.  In contrast, the regions of NGC 2068 and LBS
23N, and, to a lesser extent, NGC 2071, all show a flat distribution
out to $\sim 6$\%, where they then decline.}
\label{p4:Pcompare}
\end{figure}

\begin{figure}
\vspace*{12cm}
\includegraphics{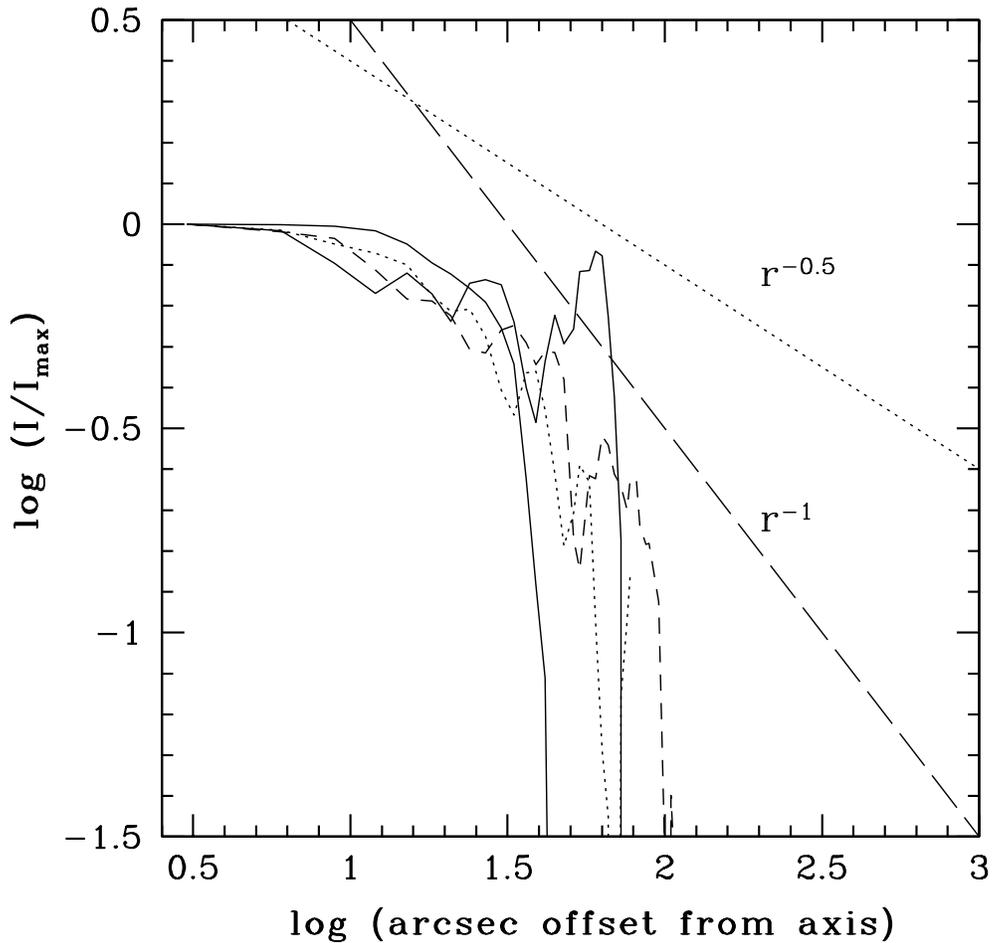}
\caption{Radial flux profile of NGC 2068. A
series of crosscuts shows the radial distribution of flux off the
filament axis.  The filament is very narrow so flux does not extend
very far off the filament's axis (the intensity peak).  Slices were
taken between cores to avoid probing the profiles of very high density
gas and dust.  The positions of the cuts are discussed in the text.
Assuming uncertainties of 10\% in the intensities measurements, the
uncertainty in each value of log$(I/I_{max})$ is $\sim 0.08$.  Slopes
of $r^{-1}$ and $r^{-0.5}$ flux profiles are shown.}
\label{p4:slice}
\end{figure}

Table \ref{p4:allthedata} contains the polarization data plotted in
Figure \ref{p4:map}.  Positional offsets are given from the J2000
coordinates $\alpha = 05^{\rm h}46^{\rm m}$33\fs8 and $\delta =
00$\degr01\arcmin43\farcs0 ($\alpha = 05^{\rm h}44^{\rm m}$00\fs0 and
$\delta = 00$\degr00\arcmin00\farcs0 in B1950).  Vectors are binned to
15\arcsec\ sampling and all have $\sigma_p > 3$, $p>1$\%, and an
absolute uncertainty in polarization percentage, $dp<1$\%.  The total
intensity at the vector position exceeds 20\% of the faintest compact
peak, OriBN 35.  This minimizes the chances of systematic effects from
chopping to a reference position, as discussed in Appendix A of
Paper II.
%\citet{mwf01}.

\begin{deluxetable}{cccc}
\tablecolumns{5} 
\tablewidth{0pc} 
\tablecaption{Observing Parameters for Jiggle Mapping} 
\tablehead{ \multicolumn{2}{c}{Pointing Center} & \colhead{Chop Position Angle} &
\colhead{Number of} \\ 
\colhead{R.A. (J2000)} & \colhead{Dec. (J2000)}
& \colhead {(east of north)} & \colhead{Times
Observed} } 
\startdata 
%#4
$05^{\rm h}46^{\rm m}$26\fs5 & $+$00\degr00\arcmin21\farcs9 & 40\degr\ & 3 \\ 
%#1
$05^{\rm h}46^{\rm m}$30\fs5 & $-$00\degr01\arcmin48\farcs4 & 90\degr\ & 3 \\ 
%#3
$05^{\rm h}46^{\rm m}$37\fs6 & $+$00\degr00\arcmin33\farcs1 & 155\degr\ & 3 \\ 
%#2
$05^{\rm h}46^{\rm m}$47\fs9 & $+$00\degr01\arcmin00\farcs3 & 155\degr\ & 3 \\ 
%f
$05^{\rm h}46^{\rm m}$25\fs7 & $+$00\degr00\arcmin34\farcs9 & 40\degr\ & 17 \\ 
%d
$05^{\rm h}46^{\rm m}$28\fs7 & $-$00\degr01\arcmin55\farcs3 & 90\degr\ & 18 \\ 
%e
$05^{\rm h}46^{\rm m}$31\fs2 & $-$00\degr00\arcmin25\farcs5 & 165\degr\ & 18 \\ 
%c
$05^{\rm h}46^{\rm m}$39\fs2 & $+$00\degr00\arcmin48\farcs9 & 155\degr\ & 13 \\ 
%b
$05^{\rm h}46^{\rm m}$44\fs7 & $+$00\degr00\arcmin33\farcs5 & 155\degr\ & 12 \\ 
%a
$05^{\rm h}46^{\rm m}$50\fs3 & $+$00\degr01\arcmin23\farcs1 & 155\degr\ & 11 \\ 
\enddata
\tablecomments{The chop throw used for all observations was
150\arcsec, with the exception of the pointing centre $\alpha =
05^{\rm h}46^{\rm m}$31\fs2 and $\delta = -$00\degr00\arcmin25\farcs5,
for which the throw was 180\arcsec. }
\label{p4:observing_parameters}
\end{deluxetable}

\begin{deluxetable}{cccccccccc}
\tablecolumns{10}
\tablewidth{0pc}
\tablecaption{Polarization Percentage and Position Angle within NGC 2068}
\tablehead{
\colhead{Region} & \colhead{\# Vectors} & \multicolumn{5}{c}{Statistics} & \multicolumn{3}{c}{Gaussian Fits} \\ 
& & \colhead{$<p>$} & \colhead{Median $p$} & \colhead{s($p$)} & \colhead{$<\theta>$} & \colhead{s($\theta$)} & \colhead{$\mu$} & \colhead{$w(\theta)$} & \colhead{$\chi^2_{red}$} \\
& & \colhead{(\%)} & \colhead{(\%)} & \colhead{(\%)} & \colhead{($^\circ$)} & \colhead{($^\circ$)} & \colhead{($^\circ$)} & \colhead{($^\circ$)} }
\startdata
1 & 68 & 4.9 & 3.6 & 3.6 & 109.2 & 41 & 123 & 24 & 0.7 \\
2 & 26 & 4.7 & 5.2 & 2.2 & 93.4 & 16 & 93 & 16 & 0.4 \\
3 & 69 & 5.2 & 4.8 & 2.9 & 74.3 & 35 & 66 & 18 & 0.7 \\
all & 163 & 5.0 & 4.3 & 3.1 & 91.4 & 40 & 95 & 41 & 0.2 \\
\enddata
\label{p4:poltable}
\end{deluxetable}

\begin{deluxetable}{rrrrrrr}
\tablecolumns{7}
\tablewidth{0pc}
\tablecaption{NGC 2068 850 \micron\ Polarization Data}
\tablehead{
\colhead{$\Delta$ R.A.} & \colhead{$\Delta$ DEC.} & \colhead{$p$} & \colhead{$dp$} 
& \colhead{$\sigma_p$} & \colhead{$\theta$} & \colhead{$d\theta$} \\
\colhead{(\arcsec)} & \colhead{(\arcsec)} & \colhead{(\%)} & \colhead{(\%)} & \colhead{} & \colhead{($^\circ$)} & \colhead{($^\circ$)}}
\startdata
  $ -36.0$ & $-244.5$ &  11.22 &  0.96 &  11.7 & $ -69.3$ &  2.5 \\ 
  $ -36.0$ & $-229.5$ &   1.65 &  0.28 &   5.8 & $ -48.9$ &  4.9 \\ 
  $ -51.0$ & $-229.5$ &   2.39 &  0.47 &   5.1 & $  51.9$ &  5.6 \\ 
  $  -6.0$ & $-214.5$ &   8.62 &  0.85 &  10.1 & $  45.0$ &  2.8 \\ 
  $ -51.0$ & $-214.5$ &   1.24 &  0.20 &   6.2 & $ -76.1$ &  4.6 \\ 
  $ -66.0$ & $-214.5$ &   5.17 &  0.59 &   8.7 & $ -82.6$ &  3.3 \\ 
  $ -21.0$ & $-199.5$ &   2.01 &  0.57 &   3.5 & $ -50.8$ &  8.1 \\ 
  $ -51.0$ & $-199.5$ &   1.96 &  0.31 &   6.2 & $  75.4$ &  4.6 \\ 
  $ -66.0$ & $-199.5$ &   3.40 &  0.44 &   7.8 & $  43.8$ &  3.7 \\ 
  $ -81.0$ & $-199.5$ &   6.16 &  0.65 &   9.5 & $  66.8$ &  3.0 \\ 
  $ -66.0$ & $-184.5$ &   6.91 &  0.41 &  16.9 & $  71.1$ &  1.7 \\ 
  $ -81.0$ & $-184.5$ &   4.46 &  0.45 &  10.0 & $ -75.7$ &  2.9 \\ 
  $ -96.0$ & $-184.5$ &   7.21 &  0.62 &  11.6 & $  46.5$ &  2.5 \\ 
  $ -66.0$ & $-169.5$ &   4.11 &  0.43 &   9.5 & $  52.5$ &  3.0 \\ 
  $ -81.0$ & $-169.5$ &   4.75 &  0.30 &  15.7 & $  73.1$ &  1.8 \\ 
  $ -96.0$ & $-169.5$ &   2.38 &  0.39 &   6.1 & $ -84.2$ &  4.7 \\ 
  $ -66.0$ & $-154.5$ &   6.16 &  0.46 &  13.5 & $  64.6$ &  2.1 \\ 
  $ -81.0$ & $-154.5$ &   2.65 &  0.27 &   9.9 & $  74.3$ &  2.9 \\ 
  $ -96.0$ & $-154.5$ &   5.59 &  0.37 &  15.2 & $  74.7$ &  1.9 \\ 
  $ -51.0$ & $-139.5$ &   4.26 &  0.52 &   8.2 & $  80.0$ &  3.5 \\ 
  $ -66.0$ & $-139.5$ &   3.10 &  0.29 &  10.5 & $  68.3$ &  2.7 \\ 
  $ -81.0$ & $-139.5$ &   3.86 &  0.25 &  15.2 & $  51.7$ &  1.9 \\ 
  $ -96.0$ & $-139.5$ &   2.28 &  0.33 &   7.0 & $   4.1$ &  4.1 \\ 
  $ -36.0$ & $-124.5$ &   6.03 &  0.43 &  14.0 & $  74.2$ &  2.0 \\ 
  $ -51.0$ & $-124.5$ &   5.70 &  0.32 &  17.6 & $  84.0$ &  1.6 \\ 
  $ -66.0$ & $-124.5$ &   3.22 &  0.23 &  14.1 & $  77.2$ &  2.0 \\ 
  $ -81.0$ & $-124.5$ &   2.86 &  0.15 &  19.6 & $  68.1$ &  1.5 \\ 
  $ -96.0$ & $-124.5$ &   2.06 &  0.19 &  10.9 & $  58.6$ &  2.6 \\ 
  $ -21.0$ & $-109.5$ &   8.99 &  0.46 &  19.5 & $  49.5$ &  1.5 \\ 
  $ -36.0$ & $-109.5$ &   6.78 &  0.59 &  11.4 & $  63.6$ &  2.5 \\ 
  $ -66.0$ & $-109.5$ &   5.12 &  0.42 &  12.2 & $  59.9$ &  2.4 \\ 
  $ -81.0$ & $-109.5$ &   1.12 &  0.15 &   7.7 & $  27.8$ &  3.7 \\ 
  $ -96.0$ & $-109.5$ &   3.22 &  0.22 &  14.5 & $  81.6$ &  2.0 \\ 
  $-126.0$ & $-109.5$ &   4.21 &  0.62 &   6.8 & $ -36.5$ &  4.2 \\ 
  $-141.0$ & $-109.5$ &   7.42 &  0.80 &   9.3 & $  40.2$ &  3.1 \\ 
  $  -6.0$ & $ -94.5$ &   6.00 &  0.63 &   9.5 & $  43.9$ &  3.0 \\ 
  $ -21.0$ & $ -94.5$ &   4.32 &  0.52 &   8.3 & $  -4.2$ &  3.5 \\ 
  $ -81.0$ & $ -94.5$ &  12.43 &  0.63 &  19.7 & $  48.2$ &  1.5 \\ 
  $ -96.0$ & $ -94.5$ &  14.80 &  0.56 &  26.5 & $  69.9$ &  1.1 \\ 
  $-111.0$ & $ -94.5$ &  11.26 &  0.74 &  15.3 & $  56.0$ &  1.9 \\ 
  $-126.0$ & $ -94.5$ &   2.80 &  0.39 &   7.2 & $  61.0$ &  4.0 \\ 
  $-141.0$ & $ -94.5$ &   4.12 &  0.34 &  12.3 & $ -87.7$ &  2.3 \\ 
  $-156.0$ & $ -94.5$ &   5.16 &  0.76 &   6.8 & $  88.2$ &  4.2 \\ 
  $   9.0$ & $ -79.5$ &   3.66 &  0.69 &   5.3 & $ -57.8$ &  5.4 \\ 
  $  -6.0$ & $ -79.5$ &   1.78 &  0.45 &   3.9 & $ -77.3$ &  7.3 \\ 
  $-126.0$ & $ -79.5$ &   4.33 &  0.49 &   8.9 & $  61.8$ &  3.2 \\ 
  $-141.0$ & $ -79.5$ &   1.18 &  0.35 &   3.4 & $ -21.7$ &  8.4 \\ 
  $-156.0$ & $ -79.5$ &   4.88 &  0.59 &   8.3 & $ -56.1$ &  3.5 \\ 
  $  24.0$ & $ -64.5$ &   5.22 &  0.67 &   7.8 & $  76.6$ &  3.7 \\ 
  $   9.0$ & $ -64.5$ &   3.30 &  0.45 &   7.4 & $  44.5$ &  3.9 \\ 
  $-111.0$ & $ -64.5$ &   6.96 &  0.71 &   9.8 & $  47.9$ &  2.9 \\ 
  $-126.0$ & $ -64.5$ &   4.77 &  0.33 &  14.3 & $  88.1$ &  2.0 \\ 
  $-141.0$ & $ -64.5$ &   5.11 &  0.28 &  18.1 & $  75.4$ &  1.6 \\ 
  $-156.0$ & $ -64.5$ &   7.09 &  0.48 &  14.9 & $  77.9$ &  1.9 \\ 
  $-171.0$ & $ -64.5$ &   7.00 &  0.91 &   7.7 & $  68.8$ &  3.7 \\ 
  $ 234.0$ & $ -49.5$ &   7.74 &  0.82 &   9.4 & $ -14.0$ &  3.0 \\ 
  $ 219.0$ & $ -49.5$ &   2.41 &  0.61 &   4.0 & $  42.0$ &  7.2 \\ 
  $ 204.0$ & $ -49.5$ &   1.42 &  0.30 &   4.8 & $  -0.1$ &  6.0 \\ 
  $ 189.0$ & $ -49.5$ &   1.72 &  0.38 &   4.6 & $  73.5$ &  6.3 \\ 
  $ 174.0$ & $ -49.5$ &   1.39 &  0.46 &   3.0 & $ -42.5$ &  9.5 \\ 
  $ 159.0$ & $ -49.5$ &   1.47 &  0.39 &   3.8 & $ -65.7$ &  7.6 \\ 
  $ 129.0$ & $ -49.5$ &   2.88 &  0.48 &   6.0 & $ -82.6$ &  4.8 \\ 
  $ 114.0$ & $ -49.5$ &   2.97 &  0.46 &   6.4 & $  71.5$ &  4.5 \\ 
  $  99.0$ & $ -49.5$ &   4.06 &  0.46 &   8.9 & $  70.8$ &  3.2 \\ 
  $  84.0$ & $ -49.5$ &   4.06 &  0.56 &   7.3 & $ -85.0$ &  3.9 \\ 
  $  69.0$ & $ -49.5$ &   4.47 &  0.67 &   6.6 & $ -71.8$ &  4.3 \\ 
  $  54.0$ & $ -49.5$ &   7.57 &  0.74 &  10.2 & $  75.6$ &  2.8 \\ 
  $  39.0$ & $ -49.5$ &   7.18 &  0.62 &  11.5 & $  81.4$ &  2.5 \\ 
  $  24.0$ & $ -49.5$ &   5.39 &  0.46 &  11.7 & $ -82.3$ &  2.4 \\ 
  $   9.0$ & $ -49.5$ &   5.09 &  0.45 &  11.3 & $ -67.5$ &  2.5 \\ 
  $  -6.0$ & $ -49.5$ &   3.51 &  0.71 &   4.9 & $  67.4$ &  5.8 \\ 
  $-111.0$ & $ -49.5$ &   6.93 &  0.62 &  11.2 & $ -12.5$ &  2.6 \\ 
  $-141.0$ & $ -49.5$ &   3.84 &  0.42 &   9.1 & $ -87.1$ &  3.1 \\ 
  $-156.0$ & $ -49.5$ &  10.14 &  0.70 &  14.4 & $ -87.9$ &  2.0 \\ 
  $ 234.0$ & $ -34.5$ &   3.63 &  0.49 &   7.4 & $   2.4$ &  3.9 \\ 
  $ 219.0$ & $ -34.5$ &   2.48 &  0.29 &   8.6 & $ -16.0$ &  3.3 \\ 
  $ 204.0$ & $ -34.5$ &   1.39 &  0.15 &   9.0 & $ -26.4$ &  3.2 \\ 
  $ 189.0$ & $ -34.5$ &   2.18 &  0.23 &   9.5 & $ -52.6$ &  3.0 \\ 
  $ 174.0$ & $ -34.5$ &   1.80 &  0.41 &   4.4 & $ -45.6$ &  6.5 \\ 
  $ 159.0$ & $ -34.5$ &   1.44 &  0.31 &   4.6 & $ -53.0$ &  6.2 \\ 
  $ 144.0$ & $ -34.5$ &   2.54 &  0.26 &   9.8 & $ -56.1$ &  2.9 \\ 
  $ 129.0$ & $ -34.5$ &   1.59 &  0.27 &   5.9 & $ -55.9$ &  4.9 \\ 
  $ 114.0$ & $ -34.5$ &   2.14 &  0.29 &   7.4 & $ -72.6$ &  3.9 \\ 
  $  99.0$ & $ -34.5$ &   5.08 &  0.26 &  19.2 & $  88.1$ &  1.5 \\ 
  $  84.0$ & $ -34.5$ &   1.43 &  0.31 &   4.7 & $  55.2$ &  6.2 \\ 
  $  69.0$ & $ -34.5$ &   3.34 &  0.27 &  12.3 & $ -72.6$ &  2.3 \\ 
  $  54.0$ & $ -34.5$ &   4.38 &  0.27 &  16.3 & $  76.5$ &  1.8 \\ 
  $  39.0$ & $ -34.5$ &   5.37 &  0.32 &  16.9 & $  81.8$ &  1.7 \\ 
  $  24.0$ & $ -34.5$ &   5.23 &  0.33 &  15.7 & $ -89.6$ &  1.8 \\ 
  $   9.0$ & $ -34.5$ &   6.04 &  0.42 &  14.3 & $  88.4$ &  2.0 \\ 
  $  -6.0$ & $ -34.5$ &   4.19 &  0.84 &   5.0 & $   7.0$ &  5.7 \\ 
  $-126.0$ & $ -34.5$ &   7.35 &  0.65 &  11.3 & $  57.2$ &  2.5 \\ 
  $-141.0$ & $ -34.5$ &  12.61 &  0.70 &  18.0 & $  65.6$ &  1.6 \\ 
  $ 249.0$ & $ -19.5$ &   4.22 &  0.86 &   4.9 & $ -36.0$ &  5.9 \\ 
  $ 234.0$ & $ -19.5$ &   5.38 &  0.57 &   9.4 & $   6.0$ &  3.0 \\ 
  $ 219.0$ & $ -19.5$ &   2.83 &  0.42 &   6.7 & $ -19.0$ &  4.3 \\ 
  $ 204.0$ & $ -19.5$ &   3.91 &  0.34 &  11.6 & $ -52.7$ &  2.5 \\ 
  $ 189.0$ & $ -19.5$ &   4.35 &  0.41 &  10.5 & $ -56.9$ &  2.7 \\ 
  $ 174.0$ & $ -19.5$ &   4.84 &  0.53 &   9.2 & $ -60.0$ &  3.1 \\ 
  $ 159.0$ & $ -19.5$ &   2.47 &  0.33 &   7.4 & $ -74.4$ &  3.9 \\ 
  $ 144.0$ & $ -19.5$ &   1.55 &  0.19 &   8.1 & $ -67.2$ &  3.5 \\ 
  $ 129.0$ & $ -19.5$ &   1.10 &  0.19 &   5.7 & $  22.6$ &  5.0 \\ 
  $  99.0$ & $ -19.5$ &   3.64 &  0.24 &  15.1 & $ -89.7$ &  1.9 \\ 
  $  84.0$ & $ -19.5$ &   2.84 &  0.31 &   9.2 & $ -82.2$ &  3.1 \\ 
  $  69.0$ & $ -19.5$ &   1.70 &  0.33 &   5.2 & $ -78.0$ &  5.6 \\ 
  $  54.0$ & $ -19.5$ &   1.94 &  0.34 &   5.8 & $  77.4$ &  5.0 \\ 
  $  39.0$ & $ -19.5$ &   9.12 &  0.39 &  23.5 & $ -83.3$ &  1.2 \\ 
  $  24.0$ & $ -19.5$ &   5.51 &  0.48 &  11.5 & $  87.3$ &  2.5 \\ 
  $   9.0$ & $ -19.5$ &   9.28 &  0.96 &   9.7 & $  90.0$ &  3.0 \\ 
  $ -96.0$ & $ -19.5$ &   5.40 &  0.52 &  10.4 & $  55.9$ &  2.7 \\ 
  $-111.0$ & $ -19.5$ &   6.52 &  0.64 &  10.3 & $  37.8$ &  2.8 \\ 
  $ 264.0$ & $  -4.5$ &   5.51 &  0.95 &   5.8 & $  39.6$ &  5.0 \\ 
  $ 249.0$ & $  -4.5$ &   4.84 &  0.54 &   8.9 & $  31.3$ &  3.2 \\ 
  $ 234.0$ & $  -4.5$ &   4.28 &  0.53 &   8.0 & $  15.5$ &  3.6 \\ 
  $ 219.0$ & $  -4.5$ &   2.82 &  0.43 &   6.6 & $ -39.5$ &  4.3 \\ 
  $ 204.0$ & $  -4.5$ &   2.89 &  0.37 &   7.7 & $ -73.0$ &  3.7 \\ 
  $ 189.0$ & $  -4.5$ &   3.55 &  0.53 &   6.7 & $ -88.2$ &  4.3 \\ 
  $ 174.0$ & $  -4.5$ &   3.93 &  0.67 &   5.9 & $ -47.0$ &  4.9 \\ 
  $ 144.0$ & $  -4.5$ &   1.53 &  0.29 &   5.2 & $  11.4$ &  5.5 \\ 
  $ 114.0$ & $  -4.5$ &   1.75 &  0.35 &   5.0 & $ -19.3$ &  5.7 \\ 
  $  99.0$ & $  -4.5$ &   3.05 &  0.25 &  12.0 & $ -88.2$ &  2.4 \\ 
  $  84.0$ & $  -4.5$ &   1.66 &  0.28 &   6.0 & $  76.1$ &  4.7 \\ 
  $  54.0$ & $  -4.5$ &   5.60 &  0.82 &   6.8 & $   8.5$ &  4.2 \\ 
  $  39.0$ & $  -4.5$ &   7.16 &  0.88 &   8.1 & $ -26.0$ &  3.5 \\ 
  $ -96.0$ & $  -4.5$ &   2.51 &  0.35 &   7.2 & $  51.9$ &  4.0 \\ 
  $-111.0$ & $  -4.5$ &   4.12 &  0.40 &  10.3 & $  27.6$ &  2.8 \\ 
  $ 249.0$ & $  10.5$ &   2.21 &  0.62 &   3.6 & $  42.6$ &  8.0 \\ 
  $ 234.0$ & $  10.5$ &   2.74 &  0.37 &   7.4 & $ -67.2$ &  3.9 \\ 
  $ 219.0$ & $  10.5$ &   3.06 &  0.26 &  11.5 & $ -77.8$ &  2.5 \\ 
  $ 204.0$ & $  10.5$ &   5.40 &  0.36 &  15.1 & $ -62.1$ &  1.9 \\ 
  $ 189.0$ & $  10.5$ &  10.42 &  0.72 &  14.4 & $ -69.2$ &  2.0 \\ 
  $ 174.0$ & $  10.5$ &   7.38 &  0.80 &   9.2 & $ -60.9$ &  3.1 \\ 
  $ 159.0$ & $  10.5$ &   5.88 &  0.73 &   8.1 & $ -46.7$ &  3.5 \\ 
  $ 144.0$ & $  10.5$ &   5.63 &  0.64 &   8.8 & $ -59.5$ &  3.3 \\ 
  $ 129.0$ & $  10.5$ &   3.20 &  0.45 &   7.1 & $ -26.9$ &  4.1 \\ 
  $ 114.0$ & $  10.5$ &   3.38 &  0.40 &   8.4 & $ -45.9$ &  3.4 \\ 
  $  99.0$ & $  10.5$ &   1.90 &  0.29 &   6.5 & $ -37.4$ &  4.4 \\ 
  $  84.0$ & $  10.5$ &   1.95 &  0.26 &   7.4 & $ -72.2$ &  3.9 \\ 
  $ -81.0$ & $  10.5$ &   5.90 &  0.66 &   9.0 & $  81.3$ &  3.2 \\ 
  $ -96.0$ & $  10.5$ &   3.76 &  0.36 &  10.5 & $  55.4$ &  2.7 \\ 
  $-111.0$ & $  10.5$ &   3.60 &  0.53 &   6.8 & $  13.4$ &  4.2 \\ 
  $ 249.0$ & $  25.5$ &  13.06 &  0.96 &  13.6 & $  -7.9$ &  2.1 \\ 
  $ 234.0$ & $  25.5$ &   6.99 &  0.53 &  13.2 & $ -51.4$ &  2.2 \\ 
  $ 219.0$ & $  25.5$ &   6.16 &  0.31 &  20.1 & $ -53.4$ &  1.4 \\ 
  $ 204.0$ & $  25.5$ &   8.99 &  0.34 &  26.1 & $ -65.6$ &  1.1 \\ 
  $ 189.0$ & $  25.5$ &  17.07 &  0.70 &  24.5 & $ -65.0$ &  1.2 \\ 
  $ 114.0$ & $  25.5$ &   8.63 &  0.86 &  10.1 & $ -33.3$ &  2.8 \\ 
  $  99.0$ & $  25.5$ &   2.83 &  0.56 &   5.0 & $ -21.0$ &  5.7 \\ 
  $  69.0$ & $  25.5$ &   5.62 &  0.92 &   6.1 & $ -53.3$ &  4.7 \\ 
  $ -81.0$ & $  25.5$ &   6.53 &  0.75 &   8.7 & $  57.9$ &  3.3 \\ 
  $ -96.0$ & $  25.5$ &   2.93 &  0.52 &   5.6 & $  64.0$ &  5.1 \\ 
  $-111.0$ & $  25.5$ &   8.23 &  0.90 &   9.1 & $ -31.6$ &  3.1 \\ 
  $ 234.0$ & $  40.5$ &   8.20 &  0.77 &  10.6 & $ -58.5$ &  2.7 \\ 
  $ 219.0$ & $  40.5$ &   8.08 &  0.43 &  18.6 & $ -64.5$ &  1.5 \\ 
  $ 204.0$ & $  40.5$ &  15.63 &  0.51 &  30.5 & $ -77.3$ &  0.9 \\ 
  $ 264.0$ & $  55.5$ &   3.27 &  0.83 &   3.9 & $ -88.3$ &  7.3 \\ 
  $ 249.0$ & $  55.5$ &   6.25 &  0.58 &  10.8 & $ -72.3$ &  2.7 \\ 
  $ 234.0$ & $  55.5$ &   6.74 &  0.47 &  14.3 & $ -49.5$ &  2.0 \\ 
  $ 219.0$ & $  55.5$ &   6.48 &  0.54 &  12.1 & $ -69.9$ &  2.4 \\ 
  $ 204.0$ & $  55.5$ &   7.45 &  0.95 &   7.8 & $ -84.8$ &  3.7 \\ 
  $ 249.0$ & $  70.5$ &   6.72 &  0.73 &   9.2 & $ -44.0$ &  3.1 \\ 
  $ 234.0$ & $  70.5$ &  12.65 &  0.59 &  21.6 & $ -59.1$ &  1.3 \\ 
  $ 219.0$ & $  70.5$ &  15.56 &  0.93 &  16.7 & $ -78.9$ &  1.7 \\ 
\enddata
\label{p4:allthedata}
\end{deluxetable}

% Figure Caption Section

\end{document}